\documentclass[a4paper,twocolumn,
english,aps,prb,floatfix,showpacs]{revtex4}
\usepackage[T1]{fontenc}
\usepackage[latin1]{inputenc}
\usepackage{amsmath}
\usepackage{babel}
\usepackage{graphics}
\usepackage{amssymb}


\begin{document}
\title
{Correlation--function distributions at the Nishimori point of two-dimensional
Ising spin glasses}
\author {S.L.A. \surname{de Queiroz}}

\email{sldq@if.ufrj.br}

\affiliation{Instituto de F\'\i sica, Universidade Federal do
Rio de Janeiro, Caixa Postal 68528, 21941-972
Rio de Janeiro RJ, Brazil}

\author {R. B. \surname{Stinchcombe}}

\email{stinch@thphys.ox.ac.uk}

\affiliation{Department of Physics, Theoretical Physics, University of
Oxford, 1 Keble Road, Oxford OX1 3NP, United Kingdom}

\date{\today}

\begin{abstract}
The multicritical behavior at the Nishimori point of two-dimensional
Ising spin glasses is investigated by using numerical transfer-matrix
methods to calculate probability distributions $P(C)$ and associated
moments of spin-spin correlation functions $C$ on strips. The angular
dependence of the shape of correlation function distributions $P(C)$
provides a stringent test of how well they obey predictions of conformal
invariance; and an even symmetry of $(1-C)\,P(C)$ reflects 
the consequences of the Ising spin-glass gauge (Nishimori) symmetry.
We show that conformal invariance is obeyed in its strictest form, and
the associated scaling of the moments of the distribution is examined,
in order to assess the validity of a recent conjecture on the exact
localization of the Nishimori point. Power law divergences
of $P(C)$ are observed near $C=1$ and $C=0$, in partial accord with a
simple scaling scheme which preserves the gauge symmetry.
\end{abstract}

\maketitle
 
\section{Introduction} 
\label{intro}
Investigation of the critical behavior of magnetic
systems is usually made harder, compared to the translationally-invariant
case, when one considers quenched disorder. This is because one gets 
inherent, randomness-induced, fluctuations in experimentally-- or
numerically measurable quantities, whose effects are often difficult to
separate from those of purely thermal origin, or connected with finite-size
scaling and crossover phenomena. 
Therefore, redoubled interest is attracted when exact results (either
rigorously proved or conjectured) are put forward in the context of
disordered spin systems. Suitable tests can then be devised in which,
by taking advantage of the proposed exact relationships, one attempts to
provide a clearer physical picture of the problem at hand.

Here we study two-dimensional $\pm J$ Ising spin glasses, i.e. 
Ising spins interacting via nearest-neighbor bonds of the same strength
and random sign. While no spin-glass ordering arises at $T \neq 0$ for
equal
concentrations of ferro-- ($p$) and antiferromagnetic ($1-p$) bonds, in
the asymmetric case $p \neq 1/2$ one can have long-range order for suitably
low concentrations of frustrated plaquettes. Accordingly, for $p$ not far
from unity a critical line on the $T-p$ plane separates paramagnetic and
ferromagnetic phases. 
  
A number of exact results have been derived along a special
line in the $T-p$ plane, 
the {\it Nishimori line}~\cite{nish81,nishbk}; also,
the exact location of a multicritical point, the {\it Nishimori point}, 
believed to be at the intersection of the
critical boundary with the Nishimori line, has been 
predicted~\cite{nn02,mnn03}.

In this paper, we investigate the multicritical behavior at
the Nishimori point. We use  numerical transfer-matrix
methods to calculate probability distributions and associated
moments of spin-spin correlation functions, and try to ascertain whether
their properties may reflect constraints imposed by conformal invariance
requirements.
 
In section~\ref{sec:review} we recall some exact statements and numerical
results, pertaining to the Nishimori line and the location of the Nishimori
point. In section \ref{sec:corfs}, we describe the numerical techniques to
be used, and provide a test of their correctness by performing
calculations on special unfrustrated systems.
In section  
\ref{sec:nrst}, we display and analyse results from
numerical calculations of correlation functions on
strips, concentrating on the shape and properties of
the corresponding distributions, especially as regards: 
(i) correlation-function equalities stemming from the
gauge symmetry obeyed by Ising spin glasses, and (ii)
conformal-invariance requirements. Section \ref{sec:appsc} is dedicated to
an approximate scaling scheme which preserves some of the essential
symmetries obeyed along the Nishimori line. Finally, in
section~\ref{sec:conc}, concluding remarks are made.

\section
{$\pm J$ Ising spin glasses}
\label{sec:review}

We consider Ising spin--$1/2$ variables on sites of a square lattice,
with couplings between nearest-neighbors $i$ and $j$ drawn from the binary
distribution:
\begin{equation}
P(J_{ij})= p\,\delta (J_{ij}-J_0)+ (1-p)\,\delta (J_{ij}+J_0)\ . 
\label{eq:1}
\end{equation}
The Nishimori line (NL) is defined by the following relationship between
temperature $T$ and positive--bond concentration $p$:
\begin{equation}
e^{-2 J_0/T} = \frac{1-p}{p}\qquad\qquad {\rm (Nishimori\ line,}\ p>
\frac{1}{2})\ .
\label{eq:2}
\end{equation}
Along this line, the configurationally-averaged internal energy is analytic
and can be calculated in closed form~\cite{nish81}. Also, a number of  
exact relationships holds
between moments of correlation function distributions, as well as
between order parameters and their derivatives.
Of particular interest here will be the following property, which has been 
shown to hold on the NL, for correlation functions between Ising 
spins $\sigma_i$, $\sigma_j$~\cite{nish81,nishbk,nish86,nish02}:
\begin{equation}
\left[ C_{ij}^{\,(2\ell +1)}\right] \equiv \left[ \langle \sigma_i \sigma_j
\rangle^{2\ell +1}\right] = \left[ C_{ij}^{\,(2\ell +2)}\right] \equiv
\left[ \langle \sigma_i \sigma_j \rangle^{2\ell +2} \right]\ ,
\label{eq:3}
\end{equation}
where angled brackets denote the usual thermal average,  square brackets
stand for configurational averages over disorder, and $\ell = 0,1, 2, 
\dots$.

It is widely believed
that, within the ordered phase, the NL separates a
high-temperature region, dominated essentially by pure-system behavior,
from one where zero--$T$ effects play the leading role. A doubly-unstable
(multicritical) fixed point is then expected to exist
on the ferro--paramagnetic phase boundary. It was argued~\cite{ldh88}
that the location of this, also known as Nishimori point (NP), should
coincide with the intersection between the critical boundary and the
NL. We shall take this view in the following.

Recently it was predicted~\cite{nn02,mnn03} that, on a square lattice,
the NP should belong to a subspace of the $T-p$ plane which is invariant
under certain duality transformations. For $\pm J$ Ising
systems, the invariant subspace is given by~\cite{nn02,mnn03}:
\begin{equation}
p\,\log (1+e^{-2J_0/T})+(1-p)\,\log (1+e^{2J_0/T})= \frac{1}{2}\,\log 2\ .
\label{eq:4}
\end{equation}
The NP is thus predicted to be at the intersection of 
Eqs.~(\ref{eq:2}) and~(\ref{eq:4}), namely $p=0.889972 \cdots\,$, 
$T/J_0 =0.956729 \cdots\,$. This agrees well (though, in some cases, it
is slightly outside estimated error bars) with available numerical
results, from which $p$ is given respectively as: $0.886(3)$ 
(series)~\cite{adler}, $0.894(2)$ (zero-$T$
calculations, assuming the phase boundary to
fall vertically from the NP)~\cite{kr97};  
$\simeq 0.885$ (exact combinatorial work)~\cite{bgp};
$0.8905(5)$ (transfer-matrix scaling of correlation lengths)~\cite{sbl3};
$0.8906(2)$ (transfer-matrix scaling of domain-wall energies)~\cite{picco};
$0.8907(2)$ (mapping into a network model for disordered
non-interacting fermions, via transfer-matrix)~\cite{mc02a};
$0.8894(9)$ (Monte-Carlo analysis of non-equilibrium 
relaxation)~\cite{ozeki}.

Though the pure-system critical point also belongs to the subspace given by
Eq.~(\ref{eq:4}), it is known~\cite{nn02} that interpreting that
equation as the exact form of the critical boundary (at least above the NP)  
brings problems of its own. Indeed, at $p=1$, the reduced slope of the
phase diagram is predicted~\cite{dom79} to be exactly:
\begin{equation}
\frac{1}{T_c}\,\frac{dT_c}{dp}\Big|_{p=1}=\frac{2 \sqrt{2}}{\ln
(\sqrt{2}+1)} = 3.2091 \cdots\ .
\label{eq:5}
\end{equation}  
While numerical transfer-matrix calculations are in good agreement with
this, giving respectively $3.25(11)$~\cite{sbl3} and 
$3.23(3)$~\cite{mc02a}, Eq.~(\ref{eq:4}) yields
$2+\sqrt{2}=3.4142
\cdots\,$, $6.4\%$ in excess of Eq.~(\ref{eq:5}). Such discrepancy is
to be compared to the disagreement between the above prediction for the NP
and the various central estimates quoted there, which never exceeds $0.5\%$
on either side. 

Thus, although the conjectured location of the  NP (to be referred 
to as CNP) given by 
Eqs.~(\ref{eq:2}) and~(\ref{eq:4}) may turn out not to be exact, it 
is certainly a good
approximation, and will be taken as the starting point in what follows.

On the other hand, conformal invariance properties~\cite{cardy} are known
to hold at the critical point of pure two-dimensional magnets, and
evidence has been provided that they are present in random systems (at
least, unfrustrated ones) as well, 
namely: random-bond Ising~\cite{dQ95,dQrbs96,dQ97}, random-bond
$q$-state Potts models~\cite{bc02}, and random transverse Ising 
chains at $T=0$ (equivalent to the two-dimensional McCoy-Wu 
model)~\cite{ir97}.
As regards the
two-dimensional Ising spin glass, early attempts to apply conformal
invariance ideas at the NP~\cite{sbl3} were undertaken in the context of
testing whether the ferro--paramagnetic transition was in the same
universality class as random percolation, as suggested by series
work~\cite{adler}. While extrapolation of the uniform zero-field
susceptibility gave the exponent ratio $\gamma/\nu=1.80(2)$, consistent
with $(\gamma/\nu)_p=43/24=1.7917 \cdots$ of percolation, the
exponent-amplitude relation~\cite{car84} $L/\pi\xi_L=\eta$ ($\xi_L$ is the
correlation length on a strip of width $L$, at criticality)
gave $\eta=0.182(5)$, thus excluding the percolation value $\eta_p=5/24
=0.208333 \cdots$. However, it must be noted that the calculated values of
$\gamma/\nu$ and $\eta$ are {\it not} mutually excludent, via the scaling
relation $\gamma/\nu=2-\eta$. This indicates that a reanalysis
of the validity of conformal invariance at the NP is in order
(though the connection to percolation can probably be ruled out,
as shown by later numerical data~\cite{picco,mc02a}). Also, one can devise
more stringent tests of conformal invariance requirements than that
given by the exponent-amplitude relation. Indeed, the correlation length
entering that relation is the single parameter governing the asymptotic
(exponential) decay of correlations, whereas for short distances
conformal invariance implies specific functional
relationships~\cite{cardy}. This fact has been exploited 
in Ref.~\onlinecite{picco},
where assorted moments of the correlation function distributions were
fitted to a form suggested by conformal invariance arguments, thus
enabling the extraction of the corresponding exponents from relatively
short-range correlation data.

Our goals here are: (1) to verify the extent to which the consequences
of Eq.~(\ref{eq:3}) are directly reflected in the probability 
distributions of the $C_{ij}$; and (2) to probe the angular dependence of
correlation functions, in order to
test whether they obey the predictions of conformal invariance.

\section{Correlation functions on strips}
\label{sec:corfs}

We apply numerical transfer-matrix (TM) methods to the spin--$1/2$ Ising 
spin glass, on strips of a square lattice of width $4 \leq L \leq 12$
sites. 
We have calculated correlation functions $C_{xy}$ between spins separated
by $x$ lattice spacings along the strip, and $y$ in the transverse
direction~\cite{dQ95}, 
such that $R=(x^2+y^2)^{1/2}$ is typically of order $L$.
On account of periodic boundary conditions across the strip, we need only 
consider $0 \leq y \leq L/2$. By iterating the TM on
sufficiently long strips, we have accumulated
enough non-overlapping samples of correlation functions (usually 
$N=10^5-10^7$), 
in order to produce clean histograms, $P(C_{xy})$, of
occurrence of $C_{xy}$. We have used a linear binning, dividing the
$[-1,1]$ interval of variation of $C_{xy}$ into $N_{\rm bin}=10^3$ equal 
bins of width
$\delta=2 \times 10^{-3}$. Furthermore, we have used a canonical ensemble
(instead of the usual grand-canonical scheme) for the bond distribution, 
so 
that overall concentration fluctuations are kept to a minimum.

Correlation functions are calculated along the lines of Section 1.4
of Ref.~\onlinecite{fs2}, with standard adaptations for an inhomogeneous 
system~\cite{dQ95}. 
For two spins in, say, row 1, separated by a distance $R$ along the strip, 
and for a given configuration ${\cal C}$ of bonds, one has:  
\begin{equation}
\langle \sigma_{0}^1 \sigma_{R}^1 \rangle_{\cal C} =
\frac{\sum_{\sigma_{0} \sigma_{R}}\tilde\psi 
(\sigma_{0})\, \sigma_{0}^1\ \left(\prod_{i=0}^{R-1} {\cal T}_{i}\right)_ 
{\sigma_{0} \sigma_{R}}\ \sigma_{R}^1\, \psi (\sigma_{R})} 
{\sum_{\sigma_{0} \sigma_{R}}  \tilde\psi (\sigma_{0})\ \left( 
\prod_{i=0}^{R-1} {\cal T}_{i}\right)_{\sigma_{0} \sigma_{R}}\ \psi 
(\sigma_{R})}\ \ ,
\label{eq:6}
\end{equation}
\noindent where $\sigma_{0} \equiv \{ \sigma_{0}^1 \ldots \sigma_{0}^L \}$ and
correspondingly for $\sigma_{R}$; the bonds that make the transfer matrices 
${\cal T}_{i}$ belong to ${\cal C}$.  For pure systems the $2^L-$component 
vectors  $\tilde\psi$, $\psi$ are determined by the boundary conditions
along the strip; 
for example, the choice of dominant left and right eigenvectors gives the 
correlation function in an infinite system\cite{fs2}. 
Here, one need only be concerned with avoiding start-up effects, since there 
is no convergence of
iterated vectors. This is done by discarding the first few hundred
iterates of the initial vectors, ${\bf v}^{T}_ 0$, ${\bf v}_ 0$. 
From then on, one can shift the dummy origin of Eq. \ref{eq:6} along the 
strip, taking $\tilde\psi$ ($\psi$) to be the
left-- (right--) iterate of ${\bf v}^{T}_ 0$ (${\bf v}_ 0$~) up to the shifted
origin, and accumulating the corresponding 
results. In order to avoid spurious correlations between the dynamical
variables, the respective iterations of $\tilde\psi$ and 
$\psi$ must use {\it distinct} realizations of the bond distribution.  

The correctness of the procedures just described can be checked
by testing them on the ferromagnetic (FM) random-bond case with a 
distribution of couplings given by
\begin{equation}
 P(J_{ij})= \frac{1}{2} ( \delta (J_{ij} -J_0) +  \delta (J_{ij} -rJ_0) ), 
\quad 0 \leq r \leq 1 \ \ {\rm (FM)}.
\label{eq:7}
\end{equation}
In this case, the exact critical temperature $\beta_c = 1/k_B T_c$ is 
known~\cite{fisch,kinzel} to be
\begin{equation}
\sinh (2\beta_{c} J_{0})\sinh (2\beta_{c}r J_{0}) = 1 \ \ {\rm (FM)}.
\label{eq:8}
\end{equation}
It is known from Monte-Carlo work~\cite{ts94} on $L \times L$ lattices,
$L \leq 1024$,
that the critical  correlation functions of the FM random-bond model are 
numerically very  close to those of a pure Ising system at its own 
criticality; for $R/L <1$ the discrepancy is always under $4\%$, and 
approaches zero as $R/L \to 0$.

\begin{figure}
{\centering \resizebox*{3.3in}{!}{\includegraphics*{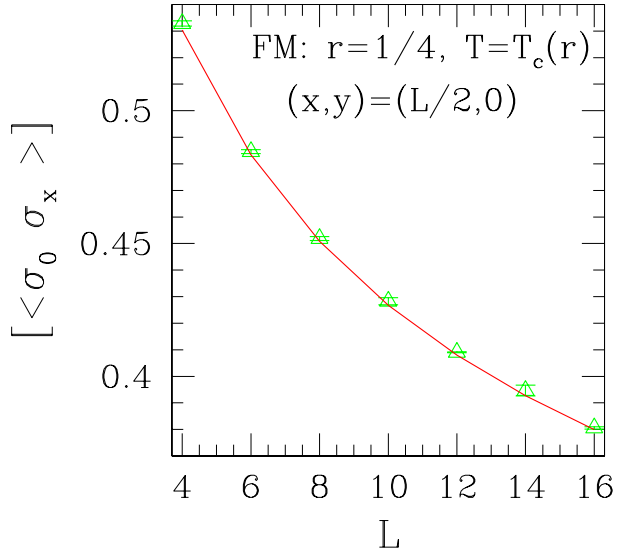}}}
\caption{
Triangles: averaged correlation functions for FM model at criticality
(see Eqs.~(\protect{\ref{eq:7}})and~(\protect{\ref{eq:8}}) ).
Error bars are twice the standard deviation among three independent runs.
Continuous line connects pure-system results for infinite strips, 
i.e. calculated  with $\tilde\psi$, $\psi$ of 
Eq.~(\protect{\ref{eq:6}}) as the dominant left and right 
eigenvectors of the TM.
} 
\label{fig:rb025}
\end{figure}

We considered a strongly disordered system with $r=1/4$, and calculated 
the averaged correlation functions $\left[ \langle\sigma_0 \sigma_R 
\rangle \right]$ (first moment of the distribution)
on strips of width $4 \leq L \leq 16$. For each strip width we made 
three independent runs, each long enough to collect $N$ non-overlapping 
samples 
of correlation functions for $x=L/2$, $y=0$ ($N=10^5$ for $L \leq 12$,
$5 \times 10^4$ and $4 \times 10^4$ respectively for $L=14$ and $16$).
The results are shown in Fig.~\ref{fig:rb025}, where error bars are
twice the standard deviation among runs~\cite{dQrbs96}. One can see
that the near identity with pure-system critical correlations is
recovered (for all points in the Figure, central estimates differ from 
their pure-system counterparts by less than $1\%$) . We 
conclude that
our calculational methods, in particular the choice of vectors
$\tilde\psi$ and $\psi$, are suitable for the description of 
correlation functions in disordered systems on strips. 

\section{Numerical results for $\pm J$ Ising spin glass on strips}
\label{sec:nrst}

\begin{figure}
{\centering \resizebox*{3.3in}{!}{\includegraphics*{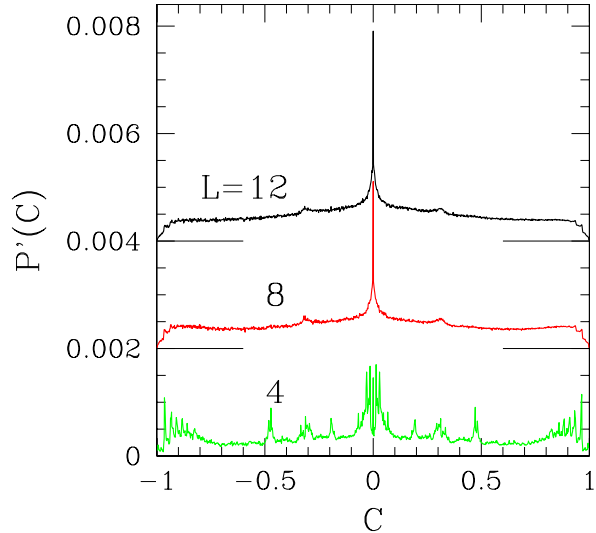}} }
\caption{
Data taken at the CNP. $P^\prime(C) \equiv (1-C)\,P(C)$, 
where $P(C)$ is the normalized 
histogram of occurrence of correlation function $C$.
Strip widths  $L=4$, $8$ and $12$ (the latter two vertically shifted
respectively  by 0.002 and 0.004, to 
avoid superposition); spin-spin relative positions ($x=L/2,y=0$). $N=2 \times 
10^6$ non-overlapping samples, for all cases. 
} 
\label{fig:cfd1}
\end{figure}

We now return to spin 
glasses. We first note that the pairing of successive odd and even moments 
predicted in Eq.~(\ref{eq:3}) implies that 
\begin{eqnarray}
\left[ C_{ij}^{\,(2\ell +1)}\right] -
\left[ C_{ij}^{\,(2\ell +2)}\right] = \hskip4.5truecm
\nonumber \\
=\int_{-1}^1 
C_{ij}^{\,(2\ell +1)}\,(1-C_{ij})\,P(C_{ij})\ dC_{ij} \equiv 0\ .
\label{eq:3a}
\end{eqnarray}
Since this holds for any $\ell$,
$P^\prime(C_{xy}) \equiv (1-C_{xy})\,P(C_{xy})$ must be an even function of
$C_{xy}$, everywhere on the NL. 
For the time being, we shall only consider $y=0$. 
Fig.~\ref{fig:cfd1}, where results at the CNP for $L=4$, $8$, 
and $12$, $(x,y)=(L/2,0)$ are displayed, illustrates that such parity 
property is present, apart from small fluctuations caused by finite-sample 
and  binning effects. The quantity $\Delta=\Delta(C) 
\equiv P^\prime(C_{xy}) -P^\prime(-C_{xy})$ provides a quantitative
measure of these deviations. With ${\overline {\Delta^k}}
\equiv [\,(N_{\rm bin}/2)^{-1}\,\sum_{C=0}^1 [\Delta 
(C)]^k\,]^{1/k}$, we find that: (i) for $N= 2 \times 10^6$,
$|{\overline {\Delta^1}}| \leq 10^{-6}$ and 
${\overline {\Delta^2}}\simeq 2 \times 10^{-5}$, constant 
against $L$; and (ii)  for fixed 
$L=4$ (where lattice structure has more pronounced effects, see
Fig.~\ref{fig:cfd1}, thus one would expect the most unfavorable
environment for measuring such small fluctuations) 
and $10^4 \leq N \leq 10^7$, ${\overline {\Delta^2}}$ varies roughly 
as $N^{-1/2}$. The latter is to be expected if deviations from parity 
arise from (randomly) incomplete sampling. 
Therefore, we can be confident that the property 
given in Eq.~(\ref{eq:3}) is satisfied by our data, except for 
small random deviations whose origin is well understood. 
For $N= 2 \times 10^6$ samples and $4 \leq L \leq 12$ the 
quantitative effect of such deviations on the moments of the 
distribution is that, for $0 \leq \ell \leq 3$, 
Eq.~(\ref{eq:3}) is satisfied to within one part in $10^4$.
This holds not only for the $y=0$ data shown above, but for all
relative positions $(x,y)$ investigated here (to be discussed
below).

As regards  conformal invariance, we first recall that, for pure
Ising systems on a strip of width $L$ with periodic boundary conditions
across, the following result holds at criticality~\cite{cardy}:
\begin{equation} 
C_{xy}^{\rm pure}\sim \left[ \frac{\pi/L}{\left( \sinh^2 (\pi
x/L)+ \sin^2 (\pi y/L)\right)^{1/2}} \right]^{\eta} \ \ ,\ \  \eta =1/4,
\label{eq:9}
\end{equation}
where the proportionality factor can be obtained from the exact
square--lattice ($L,R \to \infty,\ R \ll L$) result~\cite{wu76},
$C_{R} = 0.703\,38/R^{1/4} $. Though strictly speaking 
Eq.~(\ref{eq:9}) is an asymptotic form, we have checked that 
discrepancies are already very small at short distances: for an $L=20$
strip, the largest deviation to be found between that and
numerically-calculated values is  $1.3\%$ for $(x,y)=(1,1)$, while it
remains below $0.5\%$ for all other
cases, approaching zero very fast: for $(x,y)=(3,3)$ it is $0.2\%$.
The picture is the same already for smaller strip widths, e.g. $L=10$.

\begin{figure}
{\centering \resizebox*{3.3in}{!}{\includegraphics*{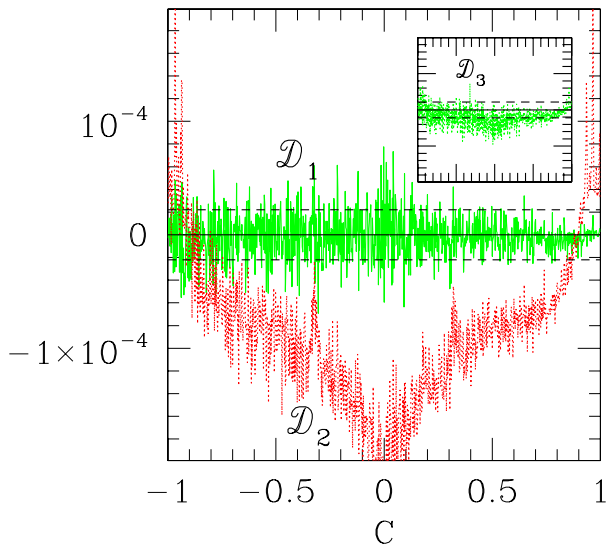}} }
\caption{
Differences ${\cal D}_i$ between symmetrized distribution functions 
$P^\prime(C_{xy})
-P^\prime(C_{x^\prime y^\prime})$
( $P^\prime(C) \equiv (1-C)\,P(C)$, where $P(C)$ is the normalized 
histogram of occurrence of correlation function $C$).
${\cal D}_1 = P^\prime (C_{15})-P^\prime (C_{23})$, ${\cal D}_2 =
P^\prime (C_{15})-P^\prime (C_{51})$; insert(same scale as main plot): 
${\cal D}_3 =P^\prime (C_{23})-P^\prime (C_{32})$.
Data taken at the CNP for strip width $L=10$, $N=2 \times 10^6$
non-overlapping samples in all cases. Dashed horizontal lines
at $\pm {\overline{\Delta^2}}$ (rms deviation of $P^\prime(C)$ from
parity for $N$ as given above, see text).
}
\label{fig:diff}
\end{figure}
Turning to disordered systems, a version of Eq.~(\ref{eq:9}) was
used at the NP in the strip calculations of Ref.~\onlinecite{picco}. 
Moments of order
$i=1, 2, \cdots 8$ of the correlation function distribution, for $y=0$ and
varying $x \leq L$, were numerically calculated and fitted to the form
Eq.~(\ref{eq:9}), with the corresponding $\eta_i$ to be extracted
from the fitting procedure. For $i=1$ their result is $\eta_1=0.1854$
(error bar estimated as $\simeq 1\%$~\cite{picco}), which agrees with, and
is more accurate than, that obtained from the asymptotic decay of
correlations, namely 0.182(5)~\cite{sbl3}.  

Here we shall try to capture variations in the correlation function 
distribution and its moments, against variations in the argument of
Eq.~(\ref{eq:9}), $z \equiv ( \sinh^2 (\pi
x/L)+ \sin^2 (\pi y/L))^{1/2}$. As (i) the coordinates
$(x,y)$ only take discrete
values, and (ii) we have no guide as to the specific functional
dependence of $P(C_{xy})$ on $z$, we shall at first compare distributions 
corresponding to points for which the argument is nearly equal,
in contrast to those corresponding e.g. to the same distance, but with
significantly different arguments. We take $L=10$ and the points
$(x,y)=(1,5)$, $(2,3)$, $(3,2)$ and $(5,1)$, for which one has
$z_{23}/z_{15}=1.001$; $z_{51}/z_{15}=2.1922$; $z_{32}/z_{23}=1.1773$.

In Fig.~\ref{fig:diff} we show the differences between pairs 
of symmetrized distributions $P^\prime (C_{xy})$: ${\cal D}_1 =
P^\prime (C_{15})-P^\prime (C_{23})$, ${\cal D}_2 =
P^\prime (C_{15})-P^\prime (C_{51})$, ${\cal D}_3 =
P^\prime (C_{23})-P^\prime (C_{32})$.
Indeed, the differences between distributions behave qualitatively
as one would expect, should the dependence on $x$ and $y$ be 
only through $z$.  The horizontal lines
at $\pm {\overline{\Delta^2}}$ (rms deviation of $P^\prime(C)$ from
parity for $N=2 \times 10^6$ as is the case here) show that, for 
${\cal D}_1$ the $0.1\%$ difference in the respective 
arguments of $z$ gives rises to effects of the same order of magnitude as 
those arising from each individual distribution's deviations from 
parity. 

\begin{figure}
{\centering \resizebox*{3.3in}{!}{\includegraphics*{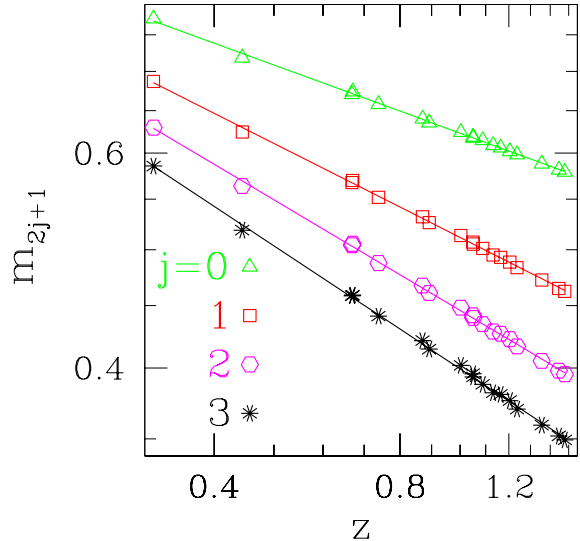}} }
\caption{
Double-logarithmic plot of odd moments of the correlation-function
distribution $P(C_{xy})$ against $z \equiv ( \sinh^2 (\pi x/L)+ \sin^2
(\pi
y/L))^{1/2}$; $1 \leq x \leq 3$, $0 \leq y \leq 5$. Straight lines are
unweighted least-squares fits to data.
Data taken at the CNP for strip width $L=10$, $N=3 \times 10^6$
non-overlapping samples in all cases. 
}
\label{fig:angdep}
\end{figure}

A more quantitative perspective can be obtained by analysing the 
dependence of assorted moments $m_i$ of the distribution against $z$.
In Fig.~\ref{fig:angdep} we show odd moments of the correlation 
function distribution at the CNP, for $L=10$ and $1 \leq x \leq 3$,
$0 \leq y \leq 5$. These correspond
to a region where the argument $z$ is strongly influenced by $y$, thus 
it is especially convenient for the discussion of the angular
dependence implied in Eq.~(\ref{eq:9}).  

From the data displayed above, the overall conclusion is that
conformal invariance holds at the NP, in the strict  angle-dependent
form corresponding to Eq.~(\ref{eq:9}). Note that the point
$(x,y)=(1,1)$, $z=1.0943$, for which the discrepancy between 
numerically-calculated 
value and asymptotic expression of correlation function is the largest for
pure systems, also corresponds to the largest deviation from least-squares 
fitting lines in the disordered case. 

\begin{figure}
{\centering \resizebox*{3.3in}{!}{\includegraphics*{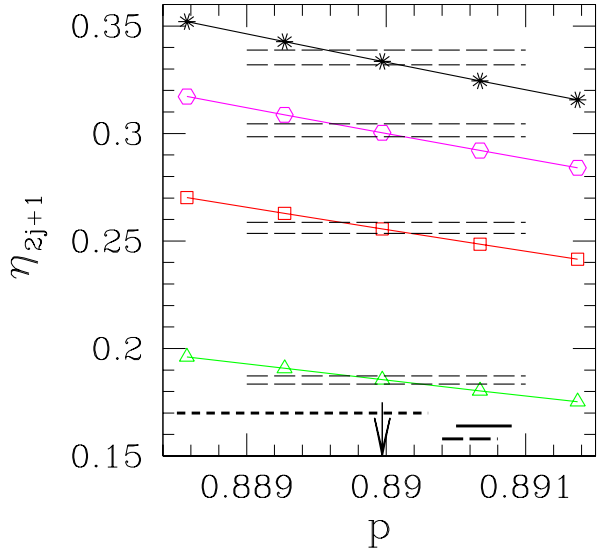}}}
\caption{
Results of least-squares fits of odd moments of the 
correlation-function distribution $P(C_{xy})$ to the form 
Eq.~(\protect{\ref{eq:9}}), with the $\eta_i$ as parameters; 
$L=10$, $1 \leq x \leq 3$, $0 \leq y \leq 5$, $N=3 \times 10^6$
non-overlapping samples in all cases. Data taken along the NL,
parametrized by ferromagnetic bond concentration $p$. Key to symbols 
is the same as in Fig.~\protect{\ref{fig:angdep}} (bottom to top:
$j=0$, $1$, $2$, $3$). Error bars are of same order as symbol sizes. 
Vertical arrow:
location of CNP. Heavy horizontal lines close to bottom show ranges of some 
recent estimates for location of NP. Long dashes: 
Ref.~\protect{\onlinecite{picco}}; full line: Ref.~\protect{\onlinecite{mc02a}};
short dashes: Ref.~\protect{\onlinecite{ozeki}}. Pairs of dashed lines
show ranges of exponent estimates given in Ref.~\protect{\onlinecite{picco}}.  
}
\label{fig:eta}
\end{figure}

We now turn to the 
question of whether the location of the NP, as conjectured 
in Refs.~\onlinecite{nn02,mnn03}, may be exact. We investigate the numerical 
values of the exponents $\eta_{\,2j+1}$, as obtained from  the moments of 
the correlation-function distribution. We have performed least-squares 
fits of data for fixed $L=10$ with the grid of $z-$values shown in
Fig.~\ref{fig:angdep} (henceforth called $z-$fits), to the form 
suggested  by Eq.~(\ref{eq:9})
for this case, namely $m_{2j+1} \sim z^{-\eta_{\,2j+1}}$. We have scanned
the NL in the immediate neighborhood of the CNP, $0.8886 \leq p \leq 
0.8914$, so as to include the 
uncertainty ranges of some recent estimates for the location of the 
NP~\cite{picco,mc02a,ozeki}. We found that the quality of fits, as given
by the corresponding chi-square,  does not change noticeably along that 
interval. Our estimates are displayed in Fig.~\ref{fig:eta}, where the 
pairs of dashed lines in the central region of the Figure show ranges of 
exponent estimates given in Ref.~\onlinecite{picco}, namely $\eta_{\,2j+1}=
0.1854$, $0.2561$, $0.3015$, $0.3354$ for $j=0$, $1$, $2$, 
$3$ (error bars assumed to be $1\%$~\cite{picco}). One sees that
the results of $z-$fits are closest to those of Ref.~\onlinecite{picco} 
at the CNP, {\it not} at the location of the NP predicted in that
reference. 

\begin{table}
\caption{\label{t1}
Estimates of exponents $\eta_{\,2j+1}$, from least-squares fits of 
averaged odd moments of 
correlation-function distributions; $z-$fit: data for $L=10$ and $1 \leq 
x \leq 3$, $0 \leq y \leq 5$, assuming  $m_{2j+1} \sim z^{-\eta_{\,2j+1}}$;
$L=4-12$: data for $x=L/2$, $y=0$, assuming $m_{\,2j+1} \sim 
L^{-\eta_{\,2j+1}}$. First and second columns: data taken at the CNP.
Third and fourth columns: pure system at criticality. 
Uncertainties in last quoted digits are shown in parentheses. 
}
\vskip 0.2cm
\begin{ruledtabular}
\begin{tabular}{@{}lllll}
& CNP & & Pure & \\
\end{tabular}
\begin{tabular}{@{}lllll}
$j$ & $z-$fit& $L=4-12$ & $z-$fit& 
exact\footnote{Given by conformal invariance~\protect{\cite{cardy}}}\\
0 & 0.1854(17)  & 0.1851(20) & 0.2497(24) & 1/4 \\
1 & 0.2556(20)  & 0.2573(32) & 0.7528(79) & 3/4\\
2 & 0.300(2)  & 0.298(4) & 1.265(15) & 5/4\\
3 & 0.334(3)  & 0.325(5) & 1.792(27) & 7/4 \\
\end{tabular}
\end{ruledtabular}
\end{table}

In order to infer how much systematic error is implicit in the $z-$fitting
procedure, we (i) performed least-squares fits of data taken at the 
CNP for $4 \leq L \leq 12$, $x=L/2$, $y=0$ against the 
form suggested by Eq.~(\ref{eq:9}) 
for that case, i.e. $m_{2j+1} \sim L^{-\eta_{\,2j+1}}$; and (ii) took
pure-system data at criticality, from the same grid of $z-$ values 
used at the NP, and extracted the corresponding estimates of $\eta_{\,2j+1}$
from $z-$ fits. The results of (i) and (ii) are shown respectively in the 
second and third columns of Table~\ref{t1}.

From the analysis of pure-system data in Table~\ref{t1} we conclude that,
at least for $j=0$ and $1$, the intrinsic inaccuracy of $z-$fits does 
not significantly compromise our final results. For these same values of
$j$, we also find greater internal consistency between $z-$ fit and
$L=4-12$ estimates at the CNP. This signals that finite-width effects
are essentially subsumed in the explicit $L-$ dependence given in
Eq~(\ref{eq:9}), i.e. higher-order finite-size corrections 
presumably do not play a significant role. Of course, a decisive test
of the latter statement would involve e.g. doing $z-$fits on
significantly wider strips.

For the moment, the data at hand allow us only to
conclude that the conjectured exact location of the NP, as given
in Refs.~\onlinecite{nn02,mnn03}, cannot be definitely ruled out. On the 
contrary, if 
we assume that the numerical values of the $\eta_i$ given in 
Ref.~\onlinecite{picco}
are the most reliable ones at present, then our results are in fact
consistent with the conjecture.
 
\section{Approximate scaling}
\label{sec:appsc}
In order to further understand the properties (so far, numerically found)
of correlation function distributions,
in this section we introduce an approximate scaling transformation which
incorporates  the gauge symmetries underlying the NL .

We recall, from real-space renormalization ideas, the concept of {\it
decimation}~\cite{sw76,ys76,rbs83}: 
considering e.g. a square lattice, summing upon the
degrees of freedom of spins on a given sublattice will produce a new
square lattice with half as many sites as the original one, and a lattice
parameter enlarged by a factor of $\sqrt{2}$. In this context, the
appropriate variable is the bond transmissivity $t \equiv \tanh
(J/T)$, which represents the
spin-spin correlation function in a restricted (single-bond) ensemble.
For a point on the NL, the constraint given in Eq.~(\ref{eq:2}) means
that, with $t_0 \equiv \tanh (J_0/T)$, the probability distribution of $t$
is
\begin{equation}
{\cal P}(t) = \frac{1}{2}(1+t_0)\,\delta (t-t_0)+ \frac{1}{
2}(1-t_0)\,\delta (t+t_0)\ .
\label{eq:app1}  
\end{equation}
 
Denoting by $t_i,\ i=1,2,3,4$ the transmissivities of bonds around a
plaquette, the scaled transmissivity of the diagonal (effective) bond after
decimation is~\cite{ys76,rbs83}      
\begin{equation}
t^\prime =\frac{t_1 t_2 + t_3 t_4}{1 + t_1 t_2 t_3 t_4}\ . 
\label{eq:app2}
\end{equation}
We investigate how the distribution function of scaled transmissivities,
${\cal P}(t^\prime)$, evolves under successive rescalings via 
Eq.~(\ref{eq:app2}), given that the original distribution obeys
Eq.~(\ref{eq:app1}).

We first prove that the distributions generated by iteration of 
Eq.~(\ref{eq:app2}) obey the NL symmetry given in 
Eq.~(\ref{eq:3}),
i.e. the NL is an invariant subspace of the transformation
Eq.~(\ref{eq:app2}). The proof is by induction, as follows.

(i) Direct examination of Eq.~(\ref{eq:app1}) shows that
the moments $m^{(0)}_j$ of the original, i.e. zero-th order distribution,
obey
\begin{equation}
m^{(0)}_{2\ell+1} \equiv \left[ t^{2\ell+1} \right]^{(0)} =
m^{(0)}_{2\ell+2}
\equiv \left[ t^{2\ell+2} \right]^{(0)}\ ,
\label{eq:app3}
\end{equation}
where the upper index $(0)$ attached to square brackets denotes average
over the distribution Eq.~(\ref{eq:app1}).

(ii) Assuming that $m^{(n)}_{2\ell+1}= m^{(n)}_{2\ell+2}$, whence
$\left[t^{2\ell+1}\,(1-t)\right]^{(n)}=0$, where the upper index $(n)$
denotes averages over the $n$--th order iterated distribution, one can show
that
\begin{eqnarray}
\left[t^{2\ell+1}\,(1-t)\right]^{(n+1)}= \hskip4.8truecm
\nonumber \\
=\left[\left(\frac
{t_1 t_2 + t_3 t_4 }{1 + t_1 t_2 t_3 t_4}\right)^{2\ell+1}\left(
1-\frac{t_1 t_2 + t_3 t_4}{1 + t_1 t_2 t_3 t_4}\right)\right]^{(n)}
\equiv 0\quad 
\label{eq:app4}  
\end{eqnarray}
by expanding the RHS and invoking the statistical independence of the
variables $t_i$. 

In fact, the invariance property just proved is guaranteed by
Eq.~(\ref{eq:app2}), because the decimated spins are connected to
an even number of bonds. Indeed, the gauge transformation originally
introduced for the $\pm 1/2$ Ising spins, in the derivation of the
NL~\cite{nish81,nish02}, could equally well have been applied to the
transmissivities of the current approach, with the same
final result.

Analytical examination of the evolution of ${\cal P}(t)$
under iteration of Eq.~(\ref{eq:app2}) shows the following
relevant features:

(1) Assuming that the distribution has a non-negligible weight near $t=0$
and making a self-consistent {\it ansatz}, in which ${\cal P}^{(n)}(t)
\sim t^{\alpha^{(n)}}$, 
one obtains the fixed-point value $\alpha^\ast=-1$, i.e. ${\cal
P}(t) \sim |t|^{-1}$, $t \to 0$.   

(2) Assuming now that the distribution has a non-negligible weight near
$t=1$, and making an {\it ansatz} in
which, with  $\delta \equiv 1-t$, ${\cal Q}^{(n)}(\delta)= {\cal
P}^{(n)}(1-t) \sim
\delta^{\beta^{(n)}}$, one obtains the fixed-point value
$\beta^\ast=-1$, i.e. ${\cal P}(t) \sim (1-t)^{-1}$, $t \to 1$.  

The analytic predictions for the exponent values $\alpha^\ast$,
$\beta^\ast$ were determined by the limiting forms of 
Eq.~(\ref{eq:app2}) for small $t_i$ or small $1-t_i$ respectively. Of
these, the first is the less reliable, since the simple transformation
Eq.~(\ref{eq:app2}) captures less well the key subtle effects needed
there, of removal of order by frustration, than the simpler characteristics
of the ferromagnetic state involved at small $1-t_i$. So we expect (and
find, see below) that the predicted $\alpha^\ast$ is much less reliable
than the predicted $\beta^\ast$.

In order to test the above predictions, one might work out the iterated
distributions directly from Eq.~(\ref{eq:app2}). This, however, is
feasible  only for the first four iterations, since the number of distinct
outcomes  grows very rapidly from 2 of Eq.~(\ref{eq:app1}) at order 
zero,
respectively to 3, 5, 25, 702 at subsequent orders
(the latter figure, for the fourth-order distribution, reflects the binning
of $25^4=390,625$ results onto $10^3$ bins of width $2 \times 
10^{-3}$,
while no binning was used for the preceding iterations). The next step
would involve considering of order $2 \times 10^{11}$ operations, which 
might be reduced by perhaps one order of magnitude by considering
symmetries. The alternative of considering a coarser binning structure (to
be kept fixed for iterations of order $\geq 4$) was seen to introduce gross
distortions of the symmetry reflected in Eq.~(\ref{eq:3}), thus the
presumptive advantages of exact recursion relations over the purely
numerical techniques of section~\ref{sec:nrst} would be lost. 
\begin{figure}
{\centering \resizebox*{3.3in}{!}{\includegraphics*{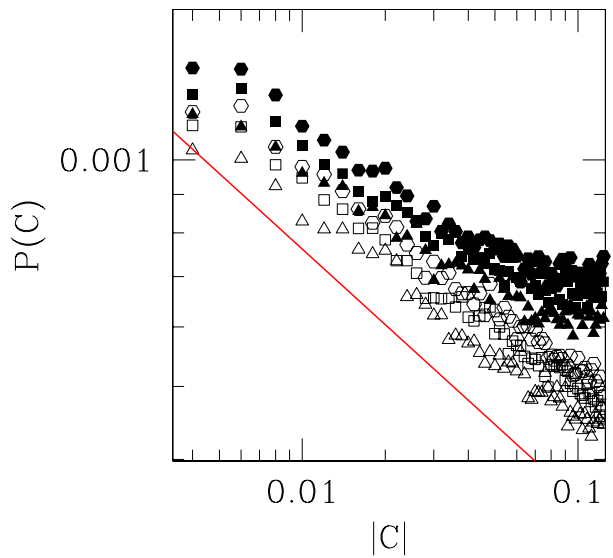}} }
\caption{Double-logarithmic plot of $P(C)$ close to $C=0$. Full
symbols: $C>0$; empty symbols: $C <0$ (divided by
$1.10$, to reduce superposition). Strip widths $L=8$ (triangles)), $10$
(squares), and $12$ (hexagons). For all cases, $x=L/2$, $y=0$.
Straight line corresponds to $P(C) \sim |C|^{-1/3}$. 
Data taken at the CNP.} 
\label{fig:ft0}
\end{figure}

\begin{figure}
{\centering \resizebox*{3.3in}{!}{\includegraphics*{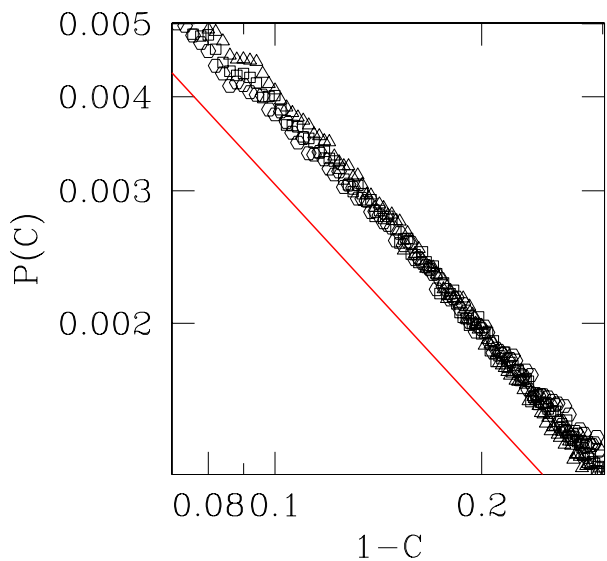}}}
\caption{Double-logarithmic plot of $P(C)$ close to $C=1$. 
Strip widths $L=8$ (triangles)), $10$ (squares), and $12$ (hexagons). For
all cases, $x=L/2$, $y=0$.
Straight line corresponds to $P(C) \sim (1-C)^{-1}$. 
Data taken at the CNP.} 
\label{fig:ft1}
\end{figure}

Before setting out to compare predictions (1) and (2) to
numerically-collected data on strips, we mention that 
the soundness of the rescaling scheme can be roughly tested as follows.
Following the usual renormalization-group ideas, one would expect the
NP to be characterized by a scaling-invariant ${\cal
P}^\ast(t)$~\cite{rbs83}. In
practice, one searches for equality between the $j$--th moments of the
$m$--th and $n$--th order iterated  distributions, expecting that the
results (i) do not depend strongly on the choice of $j$, and (ii) should
improve as $m$ and/or $n$ grows. Considering $m=0$, one can parametrize the
search for a fixed point distribution through the value $t_0$ in 
Eq.~(\ref{eq:app1}): for $t_0=t_0^\ast$, $m^{(0)}_j=m^{(n)}_j$. In 
this context, recalling the
invariance of the NL under Eq.~(\ref{eq:app2}) and taking $j=1$, 
we have $p^\ast= p(t_0^\ast)=
0.9395$ and $0.9417$, respectively for $n=3$ and $4$. 
An {\it ad hoc} two-point extrapolation
against $1/n$ gives a non-trivial limiting value, $\lim_{\ n \to \infty}
p^\ast =0.959$. This is to be compared with the presumed exact $p=0.889972
\cdots$~\cite{nn02,mnn03}.
We conclude that the rescaling scheme is qualitatively
correct, though its quantitative predictions must be carefully scrutinized.

We have tested prediction (1) above against numerical data from
strips. The results are displayed in Fig.~\ref{fig:ft0}. One can see
that, although there are clear signs of a power-law divergence both
below and above $C=0$, the corresponding exponent is certainly not above
$1/3$. The slight asymmetry between data for $C>0$ and $C<0$ is entirely
consistent with the general property that the symmetrized
distribution $P^\prime (C)$   
is an even function. Indeed, by plotting $P^\prime (C)$ we get two
essentially identical branches.  
Turning now to prediction (2), the corresponding results are displayed in
Fig.~\ref{fig:ft1}. Now one can see that the form $P(C) \sim (1-C)^{-1}$
indeed captures the essential features of behavior.

\section{Conclusions} 
\label{sec:conc}
Our investigation has probed behavior at the Nishimori point (NP), where
the infinite two-dimensional system is both critical and subject to the
Nishimori gauge symmetry. Mounting, but incomplete, evidence has suggested
that random critical two-dimensional systems share, in a statistical sense,
the conformal invariance of their pure counterparts. So it was expected 
that at the NP, the correlation function distributions $P(C)$ on strips
would be the same at all (lattice) points $(x,y)$ on any curve of constant
$|\,\sinh (\pi (x +iy)/L)\,|$ (see Eq.~(\ref{eq:9})). Evidence has
been presented in section~\ref{sec:nrst} that this strict form of
statistical conformal invariance indeed applies. This is especially
important here, since all numerical evidence available so far, regarding
conformal-invariance properties at the NP, points to values of
e.g. the exponents $\eta_i$ and the central charge which are not
obviously associated to any previously known universality 
class~\cite{sbl3,picco,mc02a} . Our data do not rule out the possibility
that the recently-conjectured location of the NP, at the intersection
of Eqs.~(\ref{eq:2}) and~(\ref{eq:4}), is indeed exact.

Among other results are the power-law divergencies of distributions $P(C)$
near $C=1$ and $C=0$, which were first identified in the invariant
distributions of the simple scaling theory (section~\ref{sec:appsc}) and
then confirmed by the strip scaling analysis (section~\ref{sec:appsc}).
Of the respective power laws ($-1$, $-1/3$) only the first was correctly
predicted because of the crudeness of the scaling transformation,
Eq.~(\ref{eq:app2}) (which nevertheless maintains the gauge 
symmetry).

\begin{acknowledgments}
We thank J T Chalker and Florian Merz for interesting discussions.
SLAdQ thanks the Department of Physics, Theoretical Physics,
at Oxford, where most of this work was carried out, for the hospitality,
and the cooperation agreement between Academia Brasileira de Ci\^encias
and the Royal Society for funding his visit. The
research of SLAdQ is partially supported by the 
Brazilian agencies CNPq (grant no 30.1692/81.5), FAPERJ (grant
nos E26--171.447/97 and E26--151.869/2000) and FUJB-UFRJ.
RBS acknowledges partial support from EPSRC Oxford Condensed Matter Theory
Programme Grant GR/R83712/01.
\end{acknowledgments}



\begin{thebibliography}{99}
\bibitem{nish81}
H. Nishimori, Prog. Theor. Phys. {\bf 66} 1169 (1981).
\bibitem{nishbk}
H. Nishimori, {\it Statistical Physics of Spin Glasses and 
Information Processing: An Introduction} (Oxford University Press, 
Oxford, 2001). 
\bibitem{nn02}
H. Nishimori and K. Nemoto, J. Phys. Soc. Jpn. {\bf 71}, 1198 (2002).
\bibitem{mnn03}
J.-M. Maillard, K. Nemoto, and H. Nishimori, cond-mat/0306154 (2003).
\bibitem{nish86}
H. Nishimori, J. Phys. Soc. Jpn. {\bf 55}, 5305 (1986).
\bibitem{nish02}
H. Nishimori, J. Phys. A {\bf 35}, 9541 (2002).
\bibitem{ldh88}
P. Le Doussal and A. B. Harris, \prl {\bf 61}, 625 (1988).
\bibitem{adler}
R. R. P. Singh  and J. Adler, \prb {\bf 54}, 364 (1996)  
\bibitem{kr97}
N. Kawashima and H. Rieger, Europhys. Lett. {\bf 39}, 85 (1997).
\bibitem{bgp}
J. A. Blackman, J. R. Gon\c calves, and J. Poulter, \pre {\bf 58} 1502 
(1998).
\bibitem{sbl3}
F. D. A. Aar\~ao Reis, S. L. A. de Queiroz, and R. R. dos Santos, 
\prb {\bf 60}, 6740 (1999).
\bibitem{picco}
A. Honecker, M. Picco, and P. Pujol, \prl {\bf 87}, 047201 (2001).
\bibitem{mc02a} 
F. Merz and J. T. Chalker, \prb {\bf 65}, 054425 (2002).
\bibitem{ozeki}
N. Ito and Y. Ozeki, Physica A {\bf 321}, 262 (2003).
\bibitem{dom79}
E. Domany, J. Phys. C {\bf 12}, L119 (1979).
\bibitem{cardy}
J. L. Cardy, in {\it Phase Transitions and Critical Phenomena},
edited by C. Domb and J. L. Lebowitz (Academic, New York, 1987), Vol. 11.
\bibitem{dQ95}
S. L. A. de Queiroz, \pre {\bf 51}, 1030 (1995).
\bibitem{dQrbs96}
S. L. A. de Queiroz and R. B. Stinchcombe, \pre {\bf 54}, 190 (1996).
\bibitem{dQ97}
S. L. A. de Queiroz, J. Phys. A {\bf 30}, L447 (1997).
\bibitem{bc02}
For a recent review, see B. Berche and C. Chatelain, cond-mat/0207421 
(lectures given at ICMP, Lviv, Ukraine).
\bibitem{ir97}
F. Igl\'oi and H. Rieger, \prl {\bf 78}, 2473 (1997).
\bibitem{car84}
J. L. Cardy, J. Phys. A {\bf 17}, L385 (1984). 
\bibitem{fs2}
M. P. Nightingale, in {\it Finite Size Scaling and Numerical 
Simulations of Statistical Systems}, edited by V. Privman (World Scientific, 
Singapore, 1990). 
\bibitem{fisch}
R. Fisch, J. Stat. Phys. {\bf 18}, 111 (1978). 
\bibitem{kinzel}
W. Kinzel and  E. Domany, \prb {\bf 23}, 3421 (1981).
\bibitem{ts94}
A. L. Talapov and L. N. Shchur, Europhys. Lett. {\bf 27}, 193 (1994).
\bibitem{wu76}
T. T. Wu, B. M. McCoy, C. A. Tracy, and E. Barouch, \prb {\bf 13}, 316 
(1976).
\bibitem{sw76}
R. B. Stinchcombe and B. P. Watson, J. Phys. C {\bf 9}, 3221 (1976).
\bibitem{ys76}
A. P. Young and R. B. Stinchcombe, J. Phys. C {\bf 9}, 4419 (1976).
\bibitem{rbs83}
R. B. Stinchcombe, in {\it Phase Transitions and Critical Phenomena},
edited by C. Domb and J. L. Lebowitz (Academic, New York, 1983), Vol. 7.
\end{thebibliography}
\end{document}